\documentclass[
reprint,
superscriptaddress,
 amsmath,amssymb,
 aps,
pre,
]{revtex4-2}

\usepackage{graphicx}
\usepackage{dcolumn}
\usepackage{bm}

\usepackage{tikz}
\definecolor{mydarkblue}{rgb}{0.0, 0.0, 1}
\usepackage[colorlinks=true,
            linkcolor=mydarkblue,
            citecolor=mydarkblue,
            urlcolor=mydarkblue]{hyperref}
\usepackage[normalem]{ulem}
\begin{document}

\preprint{}

\title{Statistical mechanics for Scrabble predicts strategy, entropy and language}

\author{Olivier Witteveen}
\affiliation{
Department of Bionanoscience, Kavli Institute of Nanoscience Delft, Technische Universiteit Delft,\\
Van der Maasweg 9, 2629 HZ Delft, the Netherlands
}

\author{Marianne Bauer}
 \email{M.S.Bauer@tudelft.nl}
\affiliation{
Department of Bionanoscience, Kavli Institute of Nanoscience Delft, Technische Universiteit Delft,\\
Van der Maasweg 9, 2629 HZ Delft, the Netherlands
}

\date{\today}

\begin{abstract}
The crossword-like patterns of tiles in Scrabble form connected graphs of occupied sites on a square lattice. 
We find the most structureless description that reproduces means and covariances observed in real Scrabble games by adapting a maximum entropy approach to connected graphs.
This pairwise model
captures the data well, and predicts word-length statistics and geometric features of the Scrabble graphs correctly; in addition, the parameters of this model are interpretable and allow us to understand Scrabble playing strategies. Using this pairwise model, we calculate entropy differences and distinguishability of Scrabble graphs across languages, without having access to the letters on the tiles. Notably, we find that the entropy is predicted better by strategic gameplay---such as word length on the board---than lexicon size. 
Finally, we find that we can use the pairwise model to correctly assign Scrabble graphs to languages, avoiding explicit feature selection and at relatively low computational cost.

\end{abstract}

\maketitle

\section{Introduction}

What does the way we play word games reveal about us and our language?
Human gameplay provides insights into characteristic behaviors, such as a player's strategy and overall performance; both can distinguish human players from artificial intelligence (AI), which now frequently outperforms humans \cite{allen2024using, AIandGames, lichtenstein_go_1980, fraenkel_computing_1981, silver_mastering_2016, collins2025evaluating}. 
In word games, both the rules and the language in which the game is played shape the space of possible outcomes, including the final configurations of the game. 
For example, the entropy of a language bounds the 
entropy of possible crossword configurations  \cite{shannon_mathematical_1948, mackay_information_2003}; 
yet, for word games in general, player choices may affect these entropies.
Calculating entropies of languages \cite{stephens_statistical_2010, takahira2016entropy,bentz2017entropy, koplenig2023large,scheibner_large_2025} or inference based on finished texts---such as identifying origin or authors \cite{mccarthy2013shakespeare, basuloewenstein, missonsingh}---is a growing area of research also due to the advent of large language models (LLMs).
Here, we focus on the board game Scrabble; we infer game rules, player types, entropy, and language from final configurations of Scrabble games with a general and interpretable maximum entropy approach.

Scrabble is a board game in the family of stochastic and partially observable games, where players do not know the entire state of the game.  AI tools exist that beat human players \cite{richards_amir, russell_artificial_2022}, even though the game is not solvable. Scrabble enjoys high popularity among households, as well as a large community of tournament Scrabble players. It is played in over 30 languages, and more recently also in online formats; in early 2026, the New York Times launched its own Scrabble-inspired app \textit{Crossplay} \cite{nytco2026crossplay}.
 
We analyze the crossword-like patterns formed by tiles on the Scrabble board at the end of the game, in absence of letters. 
We investigate Scrabble patterns from Scrabble tournaments \cite{cross_tables} and show that a maximum entropy model with pairwise interactions between squares on the board
is able to capture the data  well. This maximum entropy model  
allows us to identify game rules and strategies. 
To investigate Scrabble patterns from different languages, we adapt a Scrabble AI tool, Quackle \cite{olaughlin_quackle_2019}, to self-play. We fit $50{,}000$ games from five different languages with pairwise models. We find that Quackle games are lower in entropy compared to humans, and that, contrary to predictions for crosswords \cite{shannon_mathematical_1948,mackay_information_2003}, entropy does not increase with lexicon size for Scrabble; instead, the length of played words predicts entropy in final configurations better, consistent with the idea that playing more shorter words adds to stochasticity. 
Crossword patterns from different languages are distinguishable with an accuracy of ca. 70\%. 
Overall, we find pairwise models can capture fundamental structure, complexity and distinguishability of features of the board game Scrabble, which suggests that they may provide a useful, computationally efficient and interpretable complement to black-box models.

\begin{figure*}
    \includegraphics[width=\textwidth]{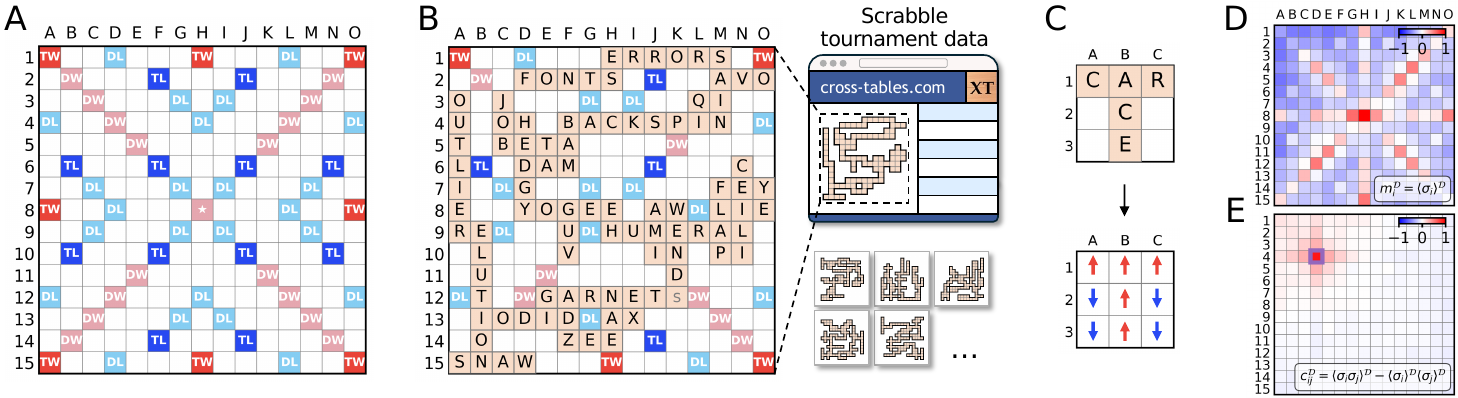}
    \caption{We model the crossword-like pattern of tiles on a Scrabble board. (A) An empty $15 {\times} 15$ Scrabble board with blank and colored, `premium' squares. DW and TW indicate double and triple word scores, respectively; DL and TL are double and triple letter scores. (B) An example of a final Scrabble board configuration: We collect ca. $50{,}000$ Scrabble games from user-uploaded and tournament data \cite{cross_tables}. All words read in top-down or left-right fashion are legal according to a standard English Scrabble lexicon \cite{naspa_nwl2023}. Lowercase, gray letters are those assigned to blank tiles by the players. (C) We assign a spin $\sigma_i \in \{-1, 1\}$ to each square on the Scrabble board to indicate whether or not a tile is present; here shown for a $3{\times}3$ example. (D) Mean square occupation from Scrabble games shows some preference for premium squares and lower occupation in top-left quadrant. 
    (E) Pairwise correlations between lattice sites, here shown for square $i = \text{D}4$ (purple box), show positive local correlations and weak anti-correlations with distant squares. }\label{fig:1}
\end{figure*}

\section{Scrabble Game}
Scrabble is a two-player word game in which players take turns at placing letter tiles on a board in a way that forms a connected pattern of words (Supp. Sec. I). Players play letter tiles from their rack of seven tiles onto the board of  $15{\times}15$ squares (Fig. \ref{fig:1}A). All tiles on the board must form words when read from left to right or top to bottom; 
all words must be included in a specific lexicon. The first word must cover the central square; words played in subsequent turns must connect to those already on the board. 

Players aim to optimize the points they earn from playing words,
with extra points awarded when they play all seven tiles  (a `bingo'). Words placed on premium squares on the board increase the number of points awarded for individual letters (DL or TL) or for the entire word (DW or TW). At the end of the game, there are $93$ to $99$ tiles on the board: in games we consider here, Scrabble ends when one of the players has finished their rack and the replenishment stock (or `bag') of initially 100 letters has run out.

We consider here the shape of the crossword-like pattern from a finished Scrabble game (Fig. \ref{fig:1}B), without letter identity. 
We assign a binary variable $\sigma_{i}\in\{-1,+1\}$ to each square $i \in \{\text{A1}, \dots, \text{O15}\}$, which denotes whether or not a square is occupied by a tile (Fig. \ref{fig:1}C). By drawing edges between adjacent tiles, $\boldsymbol{\sigma}$ implies a graph, which we refer to as a \textit{Scrabble graph}. We restrict $\boldsymbol{\sigma}$ to a domain $\mathcal{C}$ that ensures the Scrabble graph is connected and contains the permissible number of tiles.
To study the ensemble of these Scrabble graphs, we obtain ca. $50{,}000$ Scrabble graphs from the Scrabble community website Cross-Tables \cite{cross_tables}: this website has compiled data from competitive Scrabble tournaments as far back as 
1973 (Fig. \ref{fig:1}B).

\section{Maximum entropy for Scrabble}

To predict the probability of observing a particular Scrabble graph, we would need to sufficiently sample the set of all possible graphs. 
However, the configuration space is significantly larger than our ${\sim}10^5$ samples of Scrabble graphs: including disconnected patterns, there are $2^{15\times15}{\sim} 10^{67}$ configurations, and extrapolating asymptotic estimates for `polyominoes' yields ${\sim}10^{57}$ ways to connect 99 tiles together \cite{jensen_statistics_2000, jensen_enumerations_2001}.  
In order to nevertheless obtain a probability distribution $p(\boldsymbol{\sigma})$ for Scrabble graphs with our dataset, we use a maximum entropy approach:  
The Scrabble graphs we observe provide constraints on this most structureless distribution, through observables like mean square occupation and pairwise correlations, that constrain the first and second moments $\langle \sigma_i \rangle = \langle \sigma_i \rangle^{\mathcal{D}}$ and $\langle \sigma_i \sigma_j \rangle = \langle \sigma_i \sigma_j\rangle^{\mathcal{D}}$. 
Here ensemble averages $\langle \cdot \rangle$ are over configurations $\boldsymbol{\sigma} \in \mathcal{C}$
and $\langle \cdot \rangle^{\mathcal{D}}$ denotes a sample average over the data.

The means and correlations from the data that constrain $p(\boldsymbol{\sigma})$  show game-specific features (Fig. \ref{fig:1}D): players frequently place the first word horizontally, reflected by the different occupation probabilities on squares G8 and H7. The mean occupation is elevated for some premium squares, and slightly lower in the top left quadrant, likely for two reasons:  first, players may find it easier to recall words that begin with a specific letter rather than end with one. Second, letters that occur at the beginning of words statistically occur less often at end of words (e.g. `\verb+B+' starts $5.6 \%$ of words in the English Scrabble dictionary, but ends only $0.14\%$);
since the top-left quadrant of the board is more densely populated with `beginner' letters, these are then difficult to connect with from the top or from the left. Finally, we observe
positive pairwise correlations in the local neighborhood and weak anti-correlations with more distant squares on the board (see e.g. Fig. \ref{fig:1}E). 
Horizontally and vertically adjacent squares tend to be more strongly correlated than diagonal neighbors, consistent with word placement.

The most structureless distribution whose moments match observed means and pairwise correlations is the one that maximizes the entropy \cite{shannon_mathematical_1948}. It is given 
by \cite{jaynes_information_1957}
\begin{equation}\label{eq:boltzmann}
    p(\boldsymbol{\sigma}) = \frac{1}{Z} e^{-E(\boldsymbol{\sigma})},
\end{equation}
where $E(\boldsymbol{\sigma})$ is given by
\begin{equation}
    E(\boldsymbol{\sigma}) = - \sum_{i,j} J_{ij} \sigma_i \sigma_j - \sum_{i} h_i \sigma_i
\end{equation}
and the partition function $Z$ normalizes the distribution. 
In analogy with Ising spin glasses, parameters $h_i$ and $J_{ij}$ correspond to external fields and effective interactions experienced by `spin' $\sigma_i$.

Obtaining parameters $h_i$ and $J_{ij}$ from data typically involves an iterative update rule that requires repeated calculation of model expectations \cite{ackley_learning_1985}; this is tractable only for small systems \cite{nguyen_inverse_2017, lynn_exact_2025}. Instead, we use a pseudolikelihood approach, which circumvents the need to compute these expectations and requires only averages over the data \cite{aurell_inverse_2012}.
We adapted this pseudolikelihood approach to our Scrabble graphs (Methods), by restricting sample averages to the subdomain $\mathcal{F}_i \subset \mathcal{C}$
where flipping $\sigma_i$ does not break connectivity of the Scrabble graph or violate the permitted number of tiles. We derive the self-consistent equations (Methods)
\begin{align}\label{eq:pseudolikelihood_constraints1}
    \langle \sigma_i \rangle_{ \mathcal{F}_i}^{\mathcal{D}} &= \big \langle \mathrm{tanh} \, h^{\text{eff}}_i \big \rangle_{ \mathcal{F}_i}^{\mathcal{D}}, \\
    \langle \sigma_i \sigma_j \rangle_{ \mathcal{F}_i}^{\mathcal{D}} &= \big \langle \sigma_j \, \mathrm{tanh} \, h^{\text{eff}}_i \big \rangle_{ \mathcal{F}_i}^{\mathcal{D}},
    \label{eq:pseudolikelihood_constraints2}
\end{align}
where $h^{\text{eff}}_i = h_i + \sum_{j} (J_{ij} + J_{ji}) \sigma_j$ is the effective field experienced by spin $\sigma_i$, and $\langle . \rangle^{\mathcal{D}}_{\mathcal{F}_i}$ denotes the sample average over spin configurations in $\mathcal{F}_i$. 
We find the optimal parameters $h_i$ and $J_{ij}$ that solve Eqs. \ref{eq:pseudolikelihood_constraints1} and  \ref{eq:pseudolikelihood_constraints2} using a numerical gradient descent algorithm. 
In the following, we denote model matching both means and pairwise correlations with $p^{(2)}$, and a model matching only first moments (with all $J_{ij} = 0$) as $p^{(1)}$.

\section{Pairwise maximum entropy model gives insights into Scrabble}

\begin{figure*}
\includegraphics[width=0.98\textwidth]{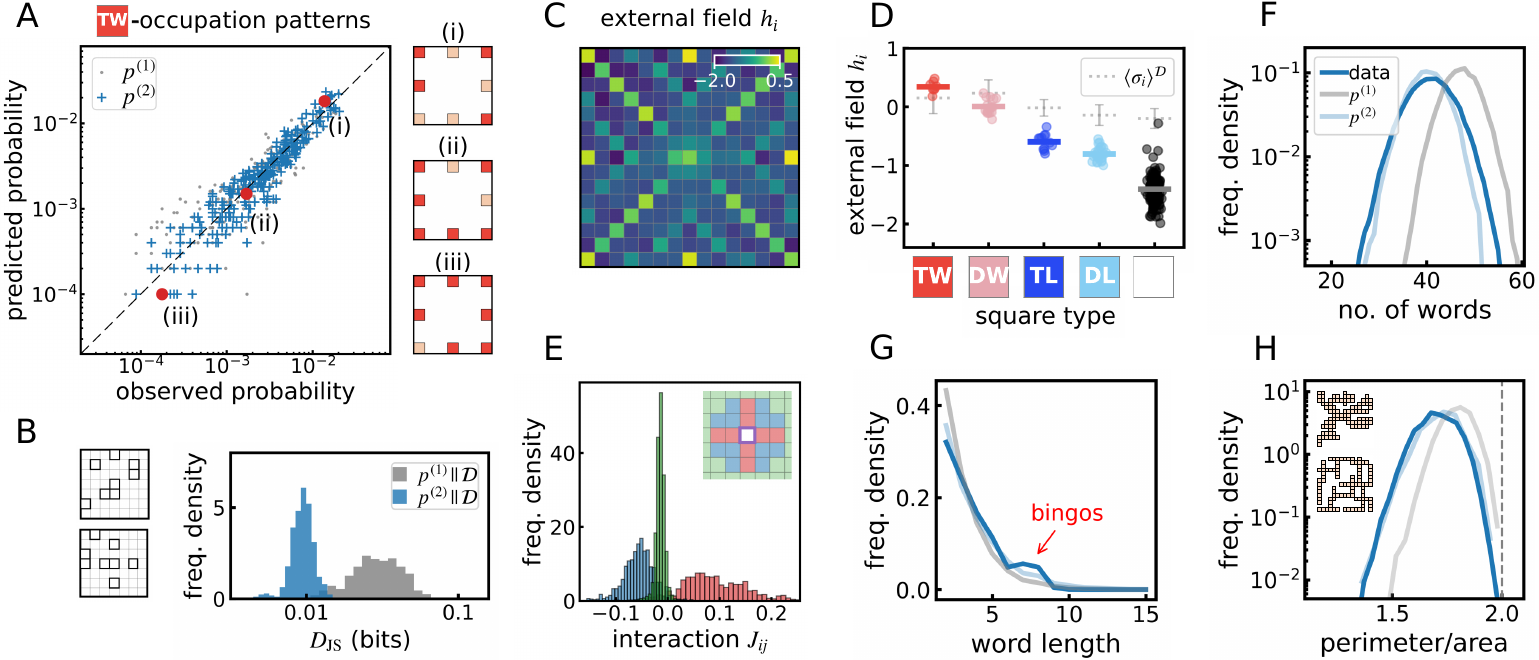}
    \caption{The pairwise maximum entropy model has interpretable parameters, and correctly predicts marginal distributions and macroscopic observables.  (A) Simultaneous occupation of the eight triple word (TW) squares on the board: for each of the $2^8 = 256$ patterns of TW-occupation, we plot the frequency of occurrence predicted by the pairwise model $p^{(2)}$ against the observed frequency (blue crosses). Red dots labeled (i-iii) are three example patterns. For comparison, the first-order model $p^{(1)}$ is shown (gray dots). (B) Jensen--Shannon divergence between observed and predicted marginal distributions for 500 random subsets of eight squares (sketch). (C) The local fields $h_i$ plotted on the Scrabble board are more symmetric than the mean occupation $m_i$ (cf. Fig. \ref{fig:1}D). (D) The external fields $h_i$ grouped by square type capture the hierarchy of premium squares: Horizontal lines are the mean $h_i$ for a given square type. The external fields capture the hierarchy better than the mean occupation (gray dashed lines; error bars are one standard deviation). (E) Histogram of interaction strengths $J_{ij}$, grouped by relative position: horizontally and vertically adjacent squares (red) show strong, positive interactions. Diagonally adjacent squares (blue) interact negatively, consistent with typical word placement. (F) Distribution of the number of words on the board at the end of a Scrabble game. (G) Distribution of word lengths on the Scrabble board; bonus points from `bingos' cause increased abundance of seven- and eight-letter words. (H) The perimeter normalized by the area:
    In the limit of an infinitely long snakelike pattern, each tile contributes two edges to the perimeter (gray broken line). Two example Scrabble graphs are shown, sampled from the left tail (top) and right tail (bottom) of the distribution, respectively.}\label{fig:2}
\end{figure*}

To show that $p^{(2)}$ describes our data effectively, we first verify that the pseudolikelihood approach is applicable to our Scrabble graphs: Means and covariances from our distribution match those from the data, with excellent agreement (Supp. Sec. II and III). 
Our model also reproduces higher-order correlations, such as  for triplets of squares (Supp. Sec. III), even though these triplets are not included as constraints.
Since a full comparison of the frequency for every Scrabble graph to data is impossible, we compare the frequency of occupation patterns for a subset of squares:  We find  an excellent match between frequencies for the set of eight triple word (TW) squares  predicted by $p^{(2)}$  and the data (Fig. \ref{fig:2}A).

Pairwise interactions are required, since $p^{(1)}$, the distribution constrained only by data means,
typically does not capture the data.
For the TW tiles (Fig. \ref{fig:2}A), 
$p^{(1)}$ works surprisingly well;  since these squares are far apart on the board,  interactions are less important. However, when
we calculate the Jensen--Shannon divergence between the observed marginal probability distributions and the models $p^{(2)}$ and $p^{(1)}$ for random subsets of squares, we
find that $p^{(2)}$ outperforms $p^{(1)}$ consistently (Fig. \ref{fig:2}B and Supp. Sec. III).

An important advantage of maximum-entropy models compared to black-box models is that they are interpretable:
Both the external fields $h_i$ and effective interactions $J_{ij}$ for the pairwise model provide insights into 
 Scrabble playing strategies (Fig. \ref{fig:2}C,D,E). 
The external fields $h_i$ reflect the bias towards adding an individual tile to the Scrabble graph. 
They display a clear a preference towards premium squares (Fig. \ref{fig:2}C, cf. Fig. \ref{fig:1}A). Compared to the mean occupation (Fig. \ref{fig:1}D), the board of $h_i$s are less affected by the directionality of the language and therefore more symmetric. Indeed, 
the external fields $h_i$ predict the hierarchy of the premium squares in Scrabble (Fig. \ref{fig:2}D): When grouped according to square type, the mean $h_i$ is highest and positive for the triple word squares (the most valuable premium square) and decreases for less valuable squares.
The mean occupation (gray) grouped by square type does not capture this trend as strongly, with more similar occupation across all square types.

Similarly, the interactions $J_{ij}$ indicate word placements more clearly than the covariances (cf. Fig. \ref{fig:1}E). The covariances are locally positive in all directions, while the $J_{ij}$ reflect the difficulties in word addition to a specific tile.
When we consider the interactions around 
a specific square (purple box, inset Fig. \ref{fig:2}E), 
interactions $J_{ij}$ further than two squares away are symmetrically distributed around an almost zero mean (green squares, inset Fig. \ref{fig:2}E). Interactions in the immediate vicinity are stronger:
We find positive $J_{ij}$s along the vertical and horizontal directions (red squares, inset Fig. \ref{fig:2}E), as words need to be read in these directions. 
In contrast, $J_{ij}$s along either diagonal are negative (blue squares, inset Fig. \ref{fig:2}E): Since the player's word has to match constraints from the vertical and horizontal words already on the board, adding words into the diagonal (in parallel  to previous words) is more difficult.
In fact, Scrabble players exploit this difficulty as a defensive tactic by constructing so-called `staircase boards' 
which are notoriously difficult to add words to.

\section{Pairwise model captures word statistics and geometry of Scrabble graphs}

To better understand Scrabble graphs and effectiveness of $p^{(2)}$, we explore Scrabble-related observables from our pairwise model $p^{(2)}$ and the data: we find that $p^{(2)}$ captures both word and tile statistics and geometric features of Scrabble graphs well.

Scrabble boards contain on average $41$ words in total at the end of the game (Fig. \ref{fig:2}F): both data and $p^{(2)}$ display a similar distribution of the number of words, which is approximately Gaussian and peaked around the mean ($41.26 \pm 0.02$ and $40.40 \pm 0.03$). The first-order model $p^{(1)}$ does not predict the number of words correctly and is shifted towards more words on the board ($47.19 \pm 0.04$). 

The mean word length on Scrabble boards is approximately 3.7 tiles. The distribution of word length is heavily skewed towards two-letter words (Fig. \ref{fig:2}G). Two-letter words are more frequent in Scrabble  because they occur as connectors between different longer words (see `\verb+RE+' connecting `\verb+OUTLIER+' and `\verb+ELUTION+' in Fig. \ref{fig:1}B)); they may also present an opportunity to play a single difficult tile (e.g. `\verb+X+' at I13 in Fig. \ref{fig:1}B, forming two two-letter words). 
The word length distribution decays for longer words, with a clear elevation at words that are seven or eight tiles long. These longer words are more frequent, because playing all seven tiles in one turn results in a `bingo', which earns more points and which players therefore strategize toward. The pairwise model $p^{(2)}$ captures the frequency of bingos better than $p^{(1)}$.
 
We probe the geometrical shape of Scrabble graphs with three observables: the perimeter (Fig. \ref{fig:2}H), the radius of gyration, and the number of holes in the pattern (Supp. Sec. V). 
We calculate a normalized perimeter by area (Fig. \ref{fig:2}H), or practically, by the number of tiles on the board, where each tile is a unit square. The distribution of normalized perimeters is bounded by $\lesssim 2$, corresponding to a connected structure where every tile contributes two on average to the perimeter; for example, by having a perfectly snakelike board where each tile has two neighbors. 
Normalized perimeters for Scrabble graphs are skewed towards this upper bound, indicative of longer words. We find an approximately normally distributed radius of gyration of approximately six units, and between zero and six holes in Scrabble graphs, all captured well by $p^{(2)}$ (Supp. Sec. V).

The impressive matches of the pairwise model to data suggest that it gives us an accurate description of English Scrabble graphs from tournaments. Now, we use this pairwise model for assessing entropy and distinguishability of Scrabble graphs from different lexica.

\section{Entropy of Scrabble graphs from different lexica}

\begin{figure*}
\includegraphics[width=0.99\textwidth]{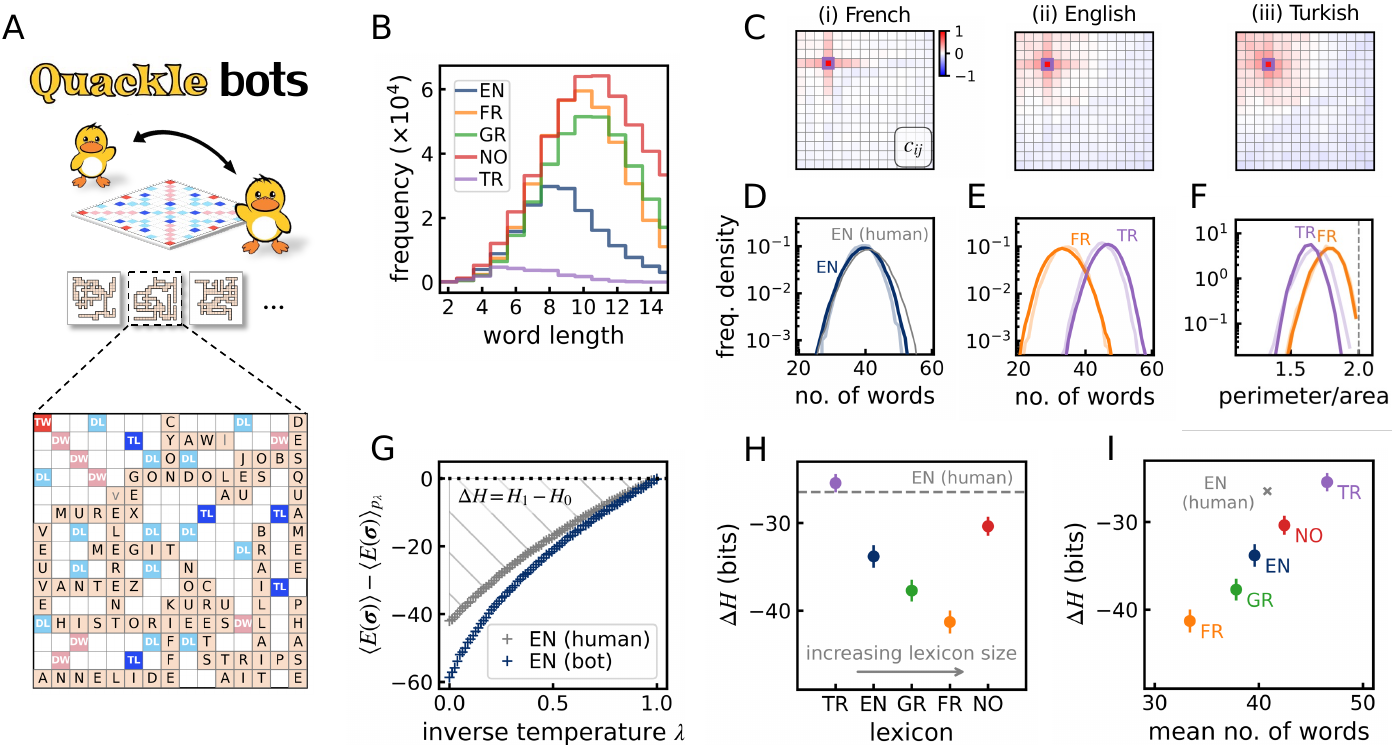}
    \caption{The size and composition of the available lexicon impacts Scrabble graphs.
    (A) We adapt a Scrabble AI tool, Quackle \cite{olaughlin_quackle_2019}, to self-play. We generate $50{,}000$ Scrabble games for each of five lexica $\ell \in \{\text{English, French, Greek, Norwegian, Turkish} \}$ included in the Quackle software distribution; an example game in French is shown. We fit a maximum entropy model $p(\boldsymbol{\sigma} \vert \ell)$ for each language. (B) The lexica differ in size and composition: Histograms showing word length frequencies in each Scrabble dictionary. (C) Languages show differences in their covariance structure. We show three examples: (i) French, (ii) English, and (iii) Turkish. (D, E) Total number of words on the board for (D) English games (both Quackle and human) and (E) French and Turkish games. (F) Normalized perimeters of French and Turkish games. In (D-F) lighter lines are predictions from the respective maximum entropy model $p(\boldsymbol{\sigma}\vert \ell)$. (G) Thermodynamic integration allows numerically stable calculation of an entropy difference relative to a reference model. The area of the shaded region corresponds to the entropy difference between human Scrabble graphs and a uniform reference model. Each data point (crosses) calculated using $2{,}500$ Monte Carlo samples of Scrabble graphs at a particular $\lambda$. English human games (gray crosses) have higher entropy than English bot games (blue crosses). (H) The entropy of the Scrabble graphs decreases with the size of the lexicon; Norwegian is an exception to this trend. Error bars are obtained by subsampling the mean energies and computing two standard deviations of the numerical integral over $\lambda$. (I) The total number of words on the board is a better predictor for the Scrabble graph entropy than the lexicon size.
    }\label{fig:3}
\end{figure*} 

To gain access to Scrabble games from different lexica to fit our pairwise model, 
we adapted a `Scrabble artificial intelligence tool', Quackle \cite{olaughlin_quackle_2019}, to self-play (Fig. \ref{fig:3}A and Supp. Sec. IV). The Quackle software distribution provides lexica for selected languages, i.e. English, French, Greek, Norwegian, and Turkish, which we study here: We generate $50{,}000$ Scrabble graphs for each lexicon. 

Both the lexicon size as well as the distribution over word lengths vary significantly across languages. Especially the Quackle lexicon for Turkish contains fewer longer words (Fig. \ref{fig:3}B); the Turkish lexicon in the Quackle distribution lists only word stems, but not words including suffixes that occur in `agglutinative' languages like Turkish. The covariance structure of Scrabble graphs indicates that word distributions from lexica are relevant (Fig. \ref{fig:3}C): English Quackle games and English human games show similar covariance matrices, while covariances in French Quackle games extend further (fewer 
and longer words) and those for Turkish display a more circular symmetry.

We fit a pairwise model $p(\boldsymbol{\sigma}\vert \ell)$ for each language $\ell$ using our adapted pseudolikelihood method, applied to the computer-generated Scrabble graphs. The pairwise model captures means and correlation structure of the Scrabble graphs successfully (Supp. Sec. V). It also correctly predicts the macroscopic observables for Scrabble graphs (Fig. \ref{fig:3}D-F and Supp. Sec. V): For example, French Scrabble graphs have fewer words and longer words on the board, whereas Scrabble graphs from our Turkish dictionary show the highest mean number of words.

\paragraph{Entropy via thermodynamic integration}

We turn to the entropy to quantify the variability of Scrabble graphs from different languages. 
Estimating this entropy requires access to a partition function, 
which becomes exponentially more difficult to estimate as system size grows. Instead, it is numerically more feasible to compute an entropy difference with respect to a non-interacting reference model. The reference entropy of this non-interacting model, $H_0$, is unknown but independent of the lexicon.
We add interactions with a (thermodynamic) integration parameter $\lambda$; then, the calculation of this entropy difference requires only the numerical calculation of average energies for each value of $\lambda$, as well as numerical integration over these values (Methods).
The integral is smooth and numerically well-behaved (Fig. \ref{fig:3}G). Formally, the entropy of the pairwise model $p^{(2)}$ is an upper bound of the true entropy of the Scrabble graph ensemble.

\paragraph{Entropy for different languages scales with the number of words on the board}

Analyses of the state space in crosswords \cite{shannon_mathematical_1948, mackay_information_2003}, where the number of possible crosswords increases with dictionary size, 
led us to expect that also here the entropy would increase with dictionary size. However, we find that almost the opposite is the case, but also not conclusively, given the stark exception of Norwegian (Fig. \ref{fig:3}H). 

Instead, the mean number of words and mean word length on the board  predict the entropy of Scrabble graphs better (Fig. \ref{fig:3}I, Supp. Fig. 4). 
We can understand this prediction by considering how the mean word number and length depends on the the rules and constraints of the Scrabble game, rather than just the lexicon size. With a hypothetically infinite dictionary, players will play all seven tiles and score a bingo in each turn, to maximize points. Even then, Scrabble graph ensembles would have non-zero entropy, as the stochasticity of tiles on the rack and previous tiles on the board affect which word connections and premium squares maximize scores. 
As the dictionary size decreases, options to play words become limited, also due to tile availability. Then, players are forced to make choices that score fewer points per turn, such as playing shorter or more disconnected words. This effective restriction increases the stochasticity in the game, which explains the increase in entropy with decreasing word length and number. The example of Norwegian illustrates that both lexicon size and tile sets matter: Norwegian has the largest lexicon, but since the tile set constrains the ease with which words can be formed, Scrabble boards display more and shorter words than e.g. French. 
  
Human Scrabble games are slightly higher in entropy than predicted from mean word number or length: notably, their entropy is higher than English Quackle games. 
Indeed, English Quackle boards show fewer words (Fig. \ref{fig:3}D) and a longer mean word length (Supp. Fig. 4) compared to human games.
This suggests that the effective lexicon practically accessible to human players is smaller than the full English dictionary accessible to Quackle, and likely increasingly depleted for longer words.
Humans tend to play shorter words but still actively optimize for bingos (Fig. \ref{fig:2}G); yet, playing longer words requires solving anagrams from the letters on the tiles. This becomes factorially more complex with the number of letters; 
as such, humans tend to be worse at this task than Quackle 
\cite{russell_artificial_2022}.

\section{Distinguishability of Scrabble graphs from different lexica}

\begin{figure*}
\includegraphics[width=0.85\textwidth]{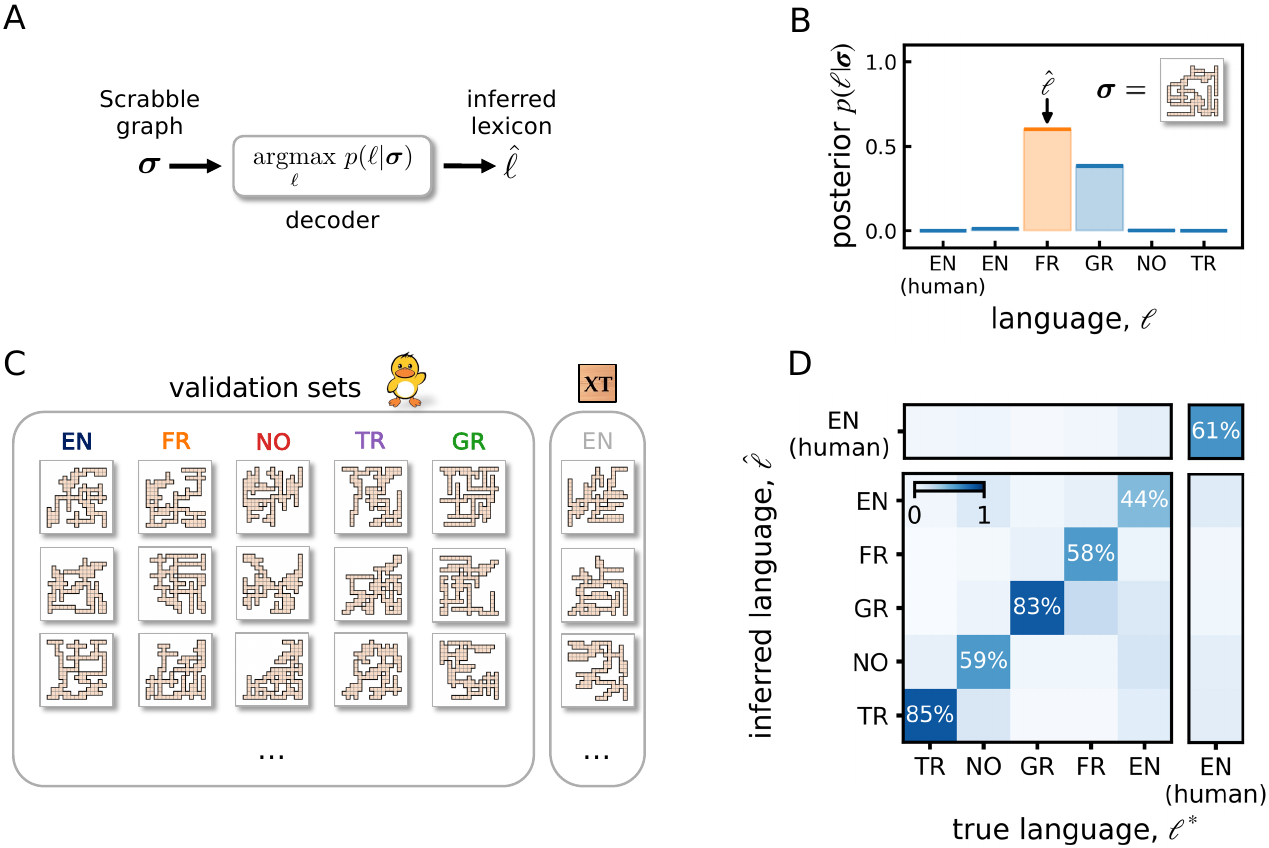}
    \caption{The pairwise maximum entropy model is successful at assigning Scrabble graphs to their corresponding lexica, despite having no access to letters. (A) 
    For a Scrabble graph $\boldsymbol{\sigma}$, the posterior distribution $p(\ell \vert \boldsymbol{\sigma})$ indicates the probability that it originated from a particular lexicon. By picking the lexicon $\hat{\ell}$ that maximizes the posterior distribution, we `decode' $\boldsymbol{\sigma}$ to obtain an estimate for the original lexicon. (B) The posterior distribution for an example Scrabble graph (inset and Fig. \ref{fig:3}A). In this case, we infer correctly that the original language is French. (C) To systematically evaluate how well we can assign Scrabble graphs to their lexica using the pairwise model, we use sets of Scrabble graphs that were not used for fitting the model. For each lexicon (Quackle and human), we use a validation set containing $2{,}000$ graphs. (D) Confusion matrix indicating the frequency at which a particular lexicon $\hat{\ell}$ was inferred, given that the true lexicon was $\ell^*$. For the set of lexica we study, we correctly infer the language $65\%$ of the time on average.}\label{fig:5}
\end{figure*}

Entropies for Scrabble graphs from different lexica showed predictable differences, but do not allow us to compare Scrabble patterns from different players (Quackle or human) or dictionaries consistently. 
To address distinguishability of different Scrabble graphs depending on lexica and players, we use an inference framework \cite{witteveen2026, Petkova_Tkacik_Bialek_Wieschaus_Gregor_2019, Bauer_Petkova_Gregor_Wieschaus_Bialek_2021} and assign each Scrabble graph $\boldsymbol{\sigma}$ a probability with which it originates from a particular dataset. 
This probability is the posterior distribution $p(\ell \vert \boldsymbol{\sigma})$, where $\ell$ indicates a lexicon. We can find this by applying Bayes' theorem, 
$p(\ell \vert \boldsymbol{\sigma}) = p(\boldsymbol{\sigma}\vert\ell) / {\sum_{\ell'} p(\boldsymbol{\sigma} \vert \ell')}$, where $p(\boldsymbol{\sigma} \vert \ell)$ 
is a pairwise model fit separately for each language $\ell$, and by assuming that all languages are equiprobable. We circumvent the calculation of the partition function as before (Methods).
For a particular graph $\boldsymbol{\sigma}$ we then find the language $\hat{\ell}$ that $\boldsymbol{\sigma}$ was most likely generated from, $\hat{\ell} = \text{argmax}_\ell \ p(\ell \vert \boldsymbol{\sigma})$ (Fig. \ref{fig:5}A, B). 
For a test set consisting of $2{,}000$ Scrabble graphs for each lexicon, we infer the language $\hat{\ell}$ that most likely generated each individual graph. Using a confusion matrix, we visualize the probability that this guess was correct (Fig. \ref{fig:5}C).  

For all languages, this inference scheme often identifies the correct language, with percentages of correctly assigned graphs ranging between 44\% (Quackle English) to 85\% (Quackle Turkish), with an average of 65\% correct assignment. These percentages would decrease if we decoded Scrabble graphs from additional languages, but nevertheless show good distinguishability. Our pairwise model was necessary for these assignments:  $p^{(1)}$, which did not capture data features correctly, fails at assessing distinguishability (average of 33\% correct assignment). Feature-based assignment, e.g based on the mean number of words, performed better than $p^{(1)}$, but still led to 
significant confusion probability
(37\% correct assignment, see Supp. Sec. VI). While additional features would improve language inference,
our approach of fitting a pairwise model that captures the data and using it for classification avoids this potentially arbitrary and non-independent feature-selection. 

Human Scrabble graphs are most similar to English Quackle graphs ($12\%$ confusion probability), even though their entropies differ by ${\sim}7$ bits, suggesting that entropy is not a good predictor of similarity. However, most word and geometric features are similar between Quackle and human English, with Quackle playing slightly longer words on average (Supp. Sec. V). 
Conversely, English Quackle graphs have the lowest probability of being identified correctly; indeed, English Quackle games frequently lie in the middle of the distribution of features (Supp. Figs. 3 and 5). They are most confusable with  Norwegian Quackle (13\%) and human Scrabble (11\%): while these percentages are similar, it is intriguing that Quackle English shows larger similarity with a different language than to human games from the same language. Since decoding based on the number of words shows the same trend, it is likely that this is due to humans selecting particular word lengths over others. 
Turkish and Greek are the most uniquely identifiable graphs, which is not captured by the word-number dependent inference, nor by inference based on $p^{(1)}$.

Overall, the high likelihood of correctly assigning languages is remarkable since our graphs contain no words; the structure of the graph alone serves as an identifier for the language that generated it.

\section{Discussion and conclusion}
Pairwise models predict data across a range of systems \cite{meshulam_statistical_2025, roudi_pairwise_2009, merchan_sufficiency_2016, cocco_inverse_2018}, with recent method developments increasing the range of possible applications \cite{carcamo_statistical_2025, kabir_overlapping_2026}. Here, the success of our pseudolikelihood-based pairwise model $p^{(2)}$ is particularly encouraging: it provides a general, computationally manageable, interpretable but not arbitrary framework to infer rules and behavioral patterns. This combination of properties allow maximum entropy models to complement other frameworks: bottom-up models on one hand, and black-box approaches on the other. Since connected structures occur in other (biological) contexts, such as polymers or biomolecules \cite{harju_multicontact_2025,loonen_phosphorylation_2025},
or tracked paths of animals or other objects \cite{bialek_statistical_2012,ahamed2021capturing, pereira_sleap_2022, nguyen_noisy_2024, valverde-mendez_macromolecular_2025,klibaite2025mapping}, our approach is broadly applicable: it could help with inference of molecule, animal or object-type without feature selection, or provide a complement to methods that require a suitable embedding \cite{marshall2021continuous,o2024dynamics}.

For Scrabble, we found that player choices (e.g. word lengths) predict entropy in final configurations better than dictionary size. This raises interesting questions on how entropy in other word puzzles might scale with the entropy of languages, analogous to predictions for crosswords \cite{shannon_mathematical_1948, mackay_information_2003}.

Our work opens up further questions on using statistical physics for understanding behavior in games.
For Scrabble, we found that we could learn about strategy and language from statistics of the final tile configurations, without considering dynamics; similarly, player type (human or AI) was identifiable from these patterns alone. 
Increased data availability could potentially
allow us to classify specific playing styles and skill levels \cite{thompson2022complex}, or identify how languages affect these styles \cite{twomey_what_2021}. 
How human or AI tendencies affect the dynamics of the game, or how they choose to adapt when they learn about the other players identity or strategy is an interesting extension of our work.

\begin{acknowledgments}
We thank William Bialek, Josh Shaevitz, Menachem Stern, Pieter Rein ten Wolde, and the members of Bauer group for useful discussions. We also thank Martin Depken for comments on the manuscript.
We acknowledge funding from the VIDI (NWO/VI.Vidi.223.169) (M.B. and O.W.). 
A special thanks to Seth Lipkin and \href{https://www.cross-tables.com}{cross-tables.com} for permission to use the data.
The authors acknowledge the use of computational resources of the DelftBlue supercomputer, provided by Delft High Performance Computing Centre. 
\end{acknowledgments}

\section*{Methods}
\subsection*{Scrabble parameter inference via pseudolikelihood}

The pairwise model in Eq. \ref{eq:boltzmann} depends on external field parameters $h_i$ and pairwise interactions $J_{ij}$. The inverse-Ising problem  to infer these parameters from data is notoriously challenging \cite{nguyen_inverse_2017}. 
Our system is different from typical spin glasses, because we are not always free to `flip' a spin $\sigma_i$ in a Scrabble graph; doing so could mean that the tiles no longer satisfy $\boldsymbol{\sigma} \in \mathcal{C}$ and form a connected crossword. Therefore, we adapt an existing method for inferring $h_i$ and $J_{ij}$ to our connected Scrabble graphs and present this method below.

To derive Eqs. \ref{eq:pseudolikelihood_constraints1}, \ref{eq:pseudolikelihood_constraints2} in the main text, we consider a subspace of Scrabble graphs $\mathcal{F}_i \subset \mathcal{C}$, where spin $\sigma_i$ can be flipped without breaking the connectivity or violating the allowed number of tiles. We define this subspace as $ \mathcal{F}_i = \big\{ \boldsymbol{\sigma} \in \mathcal{C} \, \big \vert \,  f_i[\boldsymbol{\sigma}] \in \mathcal{C} \} \subset \mathcal{C}$, where $f_i$ is an operator that flips spin $\sigma_i$ to the opposite sign.
The partition function over $\mathcal{F}_i$ is the sum of the Boltzmann weights over this subspace,
$
    Z_{\mathcal{F}_i} = \sum_{\boldsymbol{\sigma} \in \mathcal{F}_i} e^{-E(\boldsymbol{\sigma})}.
$
Since $\boldsymbol{\sigma} \in \mathcal{F}_i$, we can perform the sum over spin $\sigma_i$ separately:
\begin{equation}
    Z_{\mathcal{F}_i} = \sum_{\boldsymbol{\sigma}_{\backslash i} \in \mathcal{F}_i} 2\, \mathrm{\cosh}(h^{\text{eff}}_i) \, e^{-E_{\backslash i}(\boldsymbol{\sigma}_{\backslash i})},
\end{equation}
where $h^{\text{eff}}_i$
is the effective field experienced by spin $\sigma_i$ defined in main text and $E_{\backslash i}(\boldsymbol{\sigma}_{\backslash i}) = - \sum_{j,k \neq i} J_{jk} \sigma_j \sigma_k - \sum_{j \neq i} h_j \sigma_j$ is the Hamiltonian of the remaining spins. Writing $Z_{\mathcal{F}_i}$ in this way, it is straightforward to show:
\begin{align}
    \langle \sigma_i \rangle_{\mathcal{F}_i} &= \frac{1}{Z_{\mathcal{F}_i}}\, \frac{\partial Z_{\mathcal{F}_i}}{\partial h_i} = \big \langle \mathrm{tanh} \, h^{\text{eff}}_i \big \rangle_{\mathcal{F}_i}, \\
    \langle \sigma_i \sigma_j \rangle_{\mathcal{F}_i} &= \frac{1}{Z_{\mathcal{F}_i}} \, \frac{\partial^2 Z_{\mathcal{F}_i}}{\partial h_i \partial h_j} = \big \langle \sigma_j \, \mathrm{tanh} \, h^{\text{eff}}_i \big \rangle_{ \mathcal{F}_i},
\end{align}
where $\langle \cdot \rangle_{ \mathcal{F}_i}$ denotes the ensemble average conditional on the subdomain $\mathcal{F}_i$. The key and only approximation we make is to replace these ensemble averages with sample averages $\langle \cdot \rangle^{\mathcal{D}}_{\mathcal{F}_i}$, which yields Eqs. \ref{eq:pseudolikelihood_constraints1}, \ref{eq:pseudolikelihood_constraints2} in the main text. These equations can be solved iteratively for the Ising parameters $h_i$ and $J_{ij}$.

Equations \ref{eq:pseudolikelihood_constraints1},  \ref{eq:pseudolikelihood_constraints2} can be viewed  as conditions that maximize the \textit{pseudolikelihood} (PL) function \cite{aurell_inverse_2012}:
\begin{equation}\label{eq:plfunction}
    \mathcal{L}^{\text{PL}}(\{h_i \},\{J_{ij}\}) = \sum_{i=1}^N \bigg\langle \mathrm{log} \, p(\sigma_i \vert \boldsymbol{\sigma}_{\backslash i}, \, \boldsymbol{\sigma} \in \mathcal{F}_i) \bigg \rangle^{\mathcal{D}}_{\mathcal{F}_i},
\end{equation}
where the conditional probability is given by
$
    p(\sigma_i \vert \boldsymbol{\sigma}_{\backslash i}, \, \boldsymbol{\sigma} \in \mathcal{F}_i) = \big( 1+\sigma_i \, \mathrm{tanh} \, h^{\text{eff}}_i \big)/2.
$
Indeed, the gradients of Eq. \ref{eq:plfunction}  vanish when Eqs. \ref{eq:pseudolikelihood_constraints1}, \ref{eq:pseudolikelihood_constraints2} are satisfied:
\begin{align}\label{eq:plgrads1}
    \frac{\partial \mathcal{L}^{\text{PL}}}{\partial h_i} 
    &= \langle \sigma_i \rangle^{\mathcal{D}}_{\mathcal{F}_i} 
    - \langle \tanh h^{\text{eff}}_i \rangle^{\mathcal{D}}_{\mathcal{F}_i}, \\
    \frac{\partial \mathcal{L}^{\text{PL}}}{\partial J_{ij}} 
    &= \langle \sigma_i \sigma_j\rangle^{\mathcal{D}}_{\mathcal{F}_i} 
    - \langle \sigma_j \tanh h^{\text{eff}}_i \rangle^{\mathcal{D}}_{\mathcal{F}_i} \notag \\
    &\quad + \langle \sigma_j \sigma_i\rangle^{\mathcal{D}}_{\mathcal{F}_j} 
    - \langle \sigma_i \tanh h^{\text{eff}}_j \rangle^{\mathcal{D}}_{\mathcal{F}_j}.
    \label{eq:plgrads2}
\end{align}
To maximize the PL function in Eq. \ref{eq:plfunction}, we iteratively apply the update rules $h_i^{(\tau+1)} = h_i^{(\tau)} + \gamma_h \  \partial\mathcal{L}^{\text{PL}} / \partial h_i$ and $J_{ij}^{(\tau+1)} = J_{ij}^{(\tau)} + \gamma_J \  \partial\mathcal{L}^{\text{PL}} / \partial J_{ij}$ where $\gamma_h$ and $\gamma_J$ are learning rates. 

We note that, while Eq. \ref{eq:pseudolikelihood_constraints2} is asymmetric in $i \leftrightarrow j$, our gradient update rule for $J_{ij}$ is manifestly symmetric in this respect. By construction, we thus obtain a symmetric matrix of interactions $J_{ij}$. 

For the first-order model $p^{(1)}$, where we take only the mean spins as constraints, the external fields $h_i$ can be solved for directly using the PL method. In this case, $h_i^{\text{eff}} = h_i$, and we can directly compute $h_i = \text{arctanh}\, \langle \sigma_i \rangle_{\mathcal{F}_i}^{\mathcal{D}}$.

\subsection*{Scrabble graph entropy via thermodynamic integration} 

We calculate entropies relative to a reference 
which corresponds to a uniform, non-interacting model of Scrabble in which all Scrabble graphs are equally likely. The entropy of this uniform model, $H_0$, is independent of the lexicon, but still unknown due to the connectivity constraint of the tile pattern; a trivial upper bound is given by $H_0 \leq \log_2 2^{15\times 15}=225\text{ bits}$. 
Using a uniform model as a reference 
whose energy we set to zero, we can define a 
$\lambda$-dependent 
probability distribution $p_\lambda(\boldsymbol{\sigma}) = e^{-\lambda E(\boldsymbol{\sigma})}/Z_\lambda$, where $E(\boldsymbol{\sigma})$ is the energy according to our pairwise model $p^{(2)}$, $Z_\lambda$ the unknown partition function and $\lambda$ can tune from the uniform model ($\lambda =0$) to the pairwise model ($\lambda =1$). Instead of evaluating the partition function $Z_1 = e^{-F_1}$, we calculate the difference in free energy $\Delta F = F_1 - F_0$, 
\begin{equation}\label{eq:F}
\Delta F  = -\int_0^1 \mathrm{d} \lambda \, \frac{\mathrm{d} \, \mathrm{ln}Z_\lambda}{\mathrm{d} \lambda}  = \int_0^1 \mathrm{d} \lambda \,\langle  E(\boldsymbol{\sigma}) \rangle_{p_\lambda},
\end{equation}
where the notation  $\langle \cdot \rangle_{p_\lambda}$ indicates that the ensemble average is evaluated for a specific value of parameter $\lambda$. 
We obtain the entropy difference with respect to the reference model, $\Delta H= H_1 - H_0$, 
\begin{align}
    \Delta H  = - \Delta F + \langle E(\boldsymbol{\sigma}) \rangle
    = \int_0^1 \mathrm{d} \lambda \, (\langle E(\boldsymbol{\sigma})\rangle-\langle  E(\boldsymbol{\sigma}) \rangle_{p_\lambda}).
\end{align}
This entropy difference is computationally accessible: for a given $\lambda$, we can sample from $p_\lambda(\boldsymbol{\sigma})$ via Monte Carlo (Supp. Sec. II) to obtain the mean energy $\langle E(\boldsymbol{\sigma}) \rangle_{p_\lambda}$; the mean energy is smooth as a function of $\lambda$ and therefore its integral is numerically well-behaved (Fig. \ref{fig:3}G). 

\subsection*{Posterior distribution over languages}
To obtain the posterior distribution over languages, we take all $L$ languages as equiprobable, $p(\ell) = 1/L$,  and use Bayes' theorem 
to write the posterior distribution as:
\begin{equation}\label{eq:p2_posterior}
    p(\ell \vert \boldsymbol{\sigma}) = \frac{p(\boldsymbol{\sigma}\vert\ell)}{\sum_{\ell'} p(\boldsymbol{\sigma} \vert \ell')}= \frac{e^{\Delta F_\ell - E_\ell(\boldsymbol{\sigma})}}{\sum_{\ell'} e^{\Delta F_{\ell'} - E_{\ell'}(\boldsymbol{\sigma})}},
\end{equation}
where $p(\boldsymbol{\sigma} \vert \ell)$ is a pairwise model fit separately for each language $\ell$ and $E_\ell(\boldsymbol{\sigma})$ is its corresponding energy. 
We use the thermodynamic integral in Eq. \ref{eq:F} for each language $\ell$ to compute $\Delta F_\ell$ and avoid calculation of partition functions.

\bibliography{scrabble.bib}

\end{document}


\preprint{}

\title{Supplementary Information: Statistical mechanics for Scrabble predicts strategy, entropy and language}

\author{Olivier Witteveen}
\affiliation{
Department of Bionanoscience, Kavli Institute of Nanoscience Delft, Technische Universiteit Delft,\\
Van der Maasweg 9, 2629 HZ Delft, The Netherlands
}

\author{Marianne Bauer}
 \email{M.S.Bauer@tudelft.nl}
\affiliation{
Department of Bionanoscience, Kavli Institute of Nanoscience Delft, Technische Universiteit Delft,\\
Van der Maasweg 9, 2629 HZ Delft, The Netherlands
}

\date{\today}

\maketitle

\tableofcontents

\section{Scrabble game}

Competitive Scrabble is a two-player word game in which players are allocated letter tiles which they place onto a board in a way that forms a connected pattern of words.
Each player receives seven tiles into their rack from a set of 100 letters. In each turn, player can choose to pass, play a word, or swap up to all seven of their letters against letters from the bag. The board has a $15{\times}15$ grid of squares. On the first turn, a word must be placed so that it occupies the center square; players then take turns connecting new words to those already on the board.
All words on the board, read from left to right or top to bottom like a crossword, must be included in a specific lexicon or dictionary. When players play a word, they receive points for all new words on the board; these points depend on the score of the tiles that make up these new words, as well as the squares on the board covered by the new tiles. 

There are four types of premium squares on the board, that involve double- or triple-counting individual letter scores, DL or TL, or the scores for the entire word, DW or TW, (colored squares in Fig. 1A). In addition, players score extra points for playing all seven letters from their rack in a single turn (a `bingo'). Fundamentally, strategy in Scrabble involves maximizing the points scored during a turn, while preventing the opponent from playing valuable words or occupying premium squares in the next turn. The game ends when there are no more tiles in the bag and when one of the players has finished their rack, or when six subsequent turns have passed without a player playing a word. We ignore the latter case as it is very rare (around 2\% of the data). This means that the number of tiles on the board at the end of the game can vary between 93 and 99 in the Scrabble patterns that we consider.

\section{Monte Carlo sampling of Scrabble graphs}\label{app:mcmc}
\begin{figure}
\centering
\includegraphics[width=0.48\textwidth]{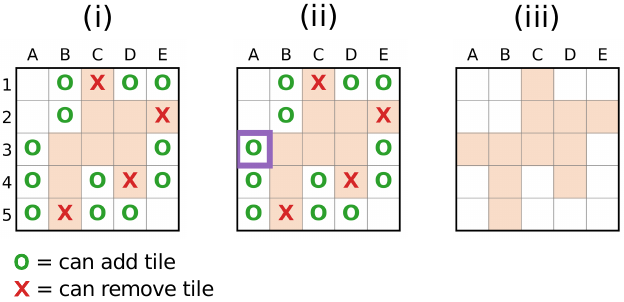}
    \caption{Monte Carlo scheme for sampling Scrabble graphs from $p(\boldsymbol{\sigma})$, consisting of three steps; we show a $5{\times5}$ example. (i) For the current configuration $\boldsymbol{\sigma}$, we find all $n(\boldsymbol{\sigma})$ places where a tile can be added (green circles) or removed (red crosses). (ii) Out of the permissible tile additions or removals, one is proposed with probability $g(\boldsymbol{\sigma} \rightarrow \boldsymbol{\sigma}') = 1/n(\boldsymbol{\sigma})$ (purple box). (iii) The proposed transition is accepted with probability $A(\boldsymbol{\sigma} \rightarrow \boldsymbol{\sigma}')$ (Eq. \ref{eq:metropolis}) or rejected. In this case, a tile addition was proposed on square A3 and accepted. By iterating (i-iii), detailed balance ensures that the Scrabble graphs are drawn from the Boltzmann distribution $p(\boldsymbol{\sigma})$. }\label{fig:mcmc}
\end{figure}

To check the predictions from the pairwise model, and to perform thermodynamic integration, we have to sample Scrabble graphs from a given distribution $p(\boldsymbol{\sigma})$. To do so, we use a Markov chain Monte Carlo scheme.
The scheme consists of three steps: First, for a given Scrabble graph, we identify all permissible tile additions and removals (e.g. green circles and red crosses in Supp. Fig. \ref{fig:mcmc}i). Equivalently, we can view these as places where a spin $\sigma_i$ can be `flipped': the number of permissible flips $n(\boldsymbol{\sigma})$ depends on the configuration $\boldsymbol{\sigma}$ itself. Second, we propose a new configuration $\boldsymbol{\sigma}'$ by randomly choosing a spin $\sigma_i$ to flip (e.g. purple box in Supp. Fig. \ref{fig:mcmc}ii). One spin is selected with uniform proposal probability $g({\boldsymbol{\sigma} \rightarrow \boldsymbol{\sigma}'}) = 1/n(\boldsymbol{\sigma})$. Third, the proposed Scrabble graph $\boldsymbol{\sigma}'$ is accepted with probability $A({\boldsymbol{\sigma} \rightarrow \boldsymbol{\sigma}'})$ or rejected (e.g. Supp. Fig. \ref{fig:mcmc}iii). 
We take the acceptance probability to be the Metropolis choice \cite{metropolis_equation_1953, hastings_monte_1970}
\begin{equation}\label{eq:metropolis}
    A(\boldsymbol{\sigma} \rightarrow \boldsymbol{\sigma}') = \mathrm{min} \bigg(1, \ \frac{n(\boldsymbol{\sigma})}{n(\boldsymbol{\sigma}')} e^{-E(\boldsymbol{\sigma}') + E(\boldsymbol{\sigma})}\bigg),
\end{equation}
such that detailed balance is satisfied. After acceptance or rejection of the new configuration, steps (i-iii) are iterated. 

\section{Pairwise maximum entropy model predicts higher-order correlations}\label{app:ho_correlations}

\begin{figure}
\centering
\includegraphics[width=0.5\textwidth]{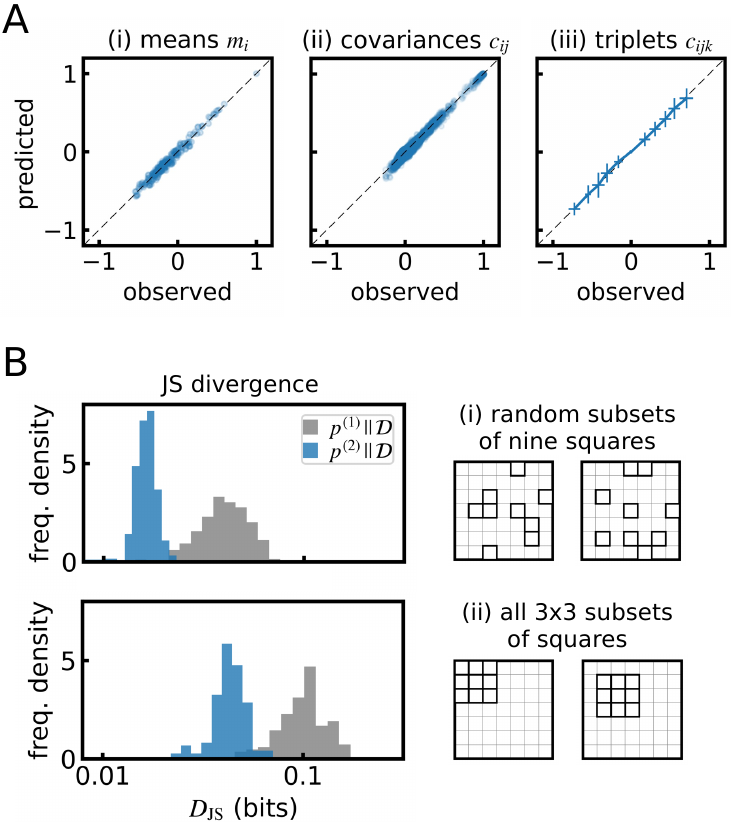}
    \caption{The pairwise maximum entropy model predicts higher-order correlations and marginal distributions. (A) Comparing first, second, and third moments from the data to those predicted by the pairwise maximum entropy model. Shown are the means $m_i = \langle \sigma_i \rangle$, pairwise correlations $c_{ij} = \langle (\sigma_i - m_i)(\sigma_j - m_j)\rangle$, and triplet correlations $c_{ijk} = \langle (\sigma_i - m_i)(\sigma_j - m_j)(\sigma_k - m_k)\rangle$. For the triplets $c_{ijk}$, instead of showing $225^3 \sim 10^7$ data points, we compare them 
    in 12 equally spaced bins. Error bars are two standard deviations of the $c_{ijk}$s in each respective bin. (B) Jensen--Shannon divergences between observed and predicted distributions for subsets of squares, corresponding to marginal distributions of $p(\boldsymbol{\sigma})$. We consider occupation of (i) 500 random subsets of nine squares and (ii) all $3{\times}3$ subsets of squares. Each subset has up to $2^9 = 512$ possible ways of being occupied by tiles. In all cases, the pairwise model $p^{(2)}$ outperforms the first-order model $p^{(1)}$ and agrees well with data. Both models perform relatively less well for the $3{\times}3$ subsets of tiles, since higher-order correlations become more important when squares are closer together.}\label{fig:supp6}
\end{figure}

To evaluate the performance of the pairwise model $p^{(2)}$, we first verify that our pseudo-likelihood (PL) method satisfies the original set of constraints $\langle \sigma_i \rangle = \langle \sigma_i \rangle^{\mathcal{D}}$ and $\langle \sigma_i \sigma_j \rangle = \langle \sigma_i \sigma_j \rangle^{\mathcal{D}}$. Indeed, we find that the model $p^{(2)}$ correctly predicts means and pairwise correlations (Supp. Fig. \ref{fig:supp6}Ai--ii). It is less trivial that a model constrained only by pairwise statistics should capture higher-order interactions between more than two squares on the Scrabble board: Nevertheless, we find that the pairwise model convincingly captures correlations between triplets of squares (Supp. Fig. \ref{fig:supp6}Aiii), similar to observations made for neuronal models \cite{schneidman_weak_2006, meshulam_statistical_2025}. 

To determine the accuracy of $p^{(2)}$ in full, one could compare the predicted probability of observing a Scrabble graph to how frequently it is observed in the data. However, due to undersampling of the full configuration space, no full Scrabble graph occurs more than once in our dataset. Instead, we compare predictions for occupation of subsets of squares $\boldsymbol{\sigma}_\text{s}$, corresponding to marginal distributions of $p(\boldsymbol{\sigma})$. In the main text, we considered the set of TW-squares as a specific example (Fig. 2A). Here, we more systematically consider (i) randomly chosen subsets of nine squares and (ii) $3{\times}3$ groups of squares (Supp. Fig. \ref{fig:supp6}Bi--ii). For each subset, we calculate the Jensen--Shannon divergence \cite{lin1991divergence, cover_elements_1991} between the predicted distribution $p(\boldsymbol{\sigma}_\text{s})$ and the observed distribution $p^\mathcal{D}(\boldsymbol{\sigma}_\text{s})$,
\begin{equation}\label{eq:jensenshannon}
     D_\text{JS}\big(p(\boldsymbol{\sigma}_\text{s}) \Vert p^{\mathcal{D}}(\boldsymbol{\sigma}_\text{s})\big) = \frac{1}{2} D_\text{KL}\big(p(\boldsymbol{\sigma}_\text{s}) \Vert m(\boldsymbol{\sigma}_\text{s})\big) + \frac{1}{2} D_\text{KL}\big(p^\mathcal{D}(\boldsymbol{\sigma}_\text{s}) \Vert  m(\boldsymbol{\sigma_\text{s}})\big),
\end{equation}
where $m(\boldsymbol{\sigma_\text{s}}) = (p(\boldsymbol{\sigma}_\text{s}) + p^{\mathcal{D}}(\boldsymbol{\sigma}_\text{s}))/2$ is a mixture distribution and $D_\text{KL}(p(x) \Vert  q(x)) = \sum_{x \in \mathcal{X}} p(x) \, \mathrm{log} \big(p(x)/q(x) \big)$
is the Kullback--Leibler divergence \cite{cover_elements_1991}. 
In all cases, the pairwise model $p^{(2)}$ agrees well with the data and performs better than model $p^{(1)}$, which is only constrained by the mean occupation of the squares. Both models $p^{(1)}$ and $p^{(2)}$ perform comparatively less well when squares are closer together, such as in $3{\times}3$ groups, since higher-order interactions between squares become more important for squares in closer proximity.

\section{Quackle simulation details}

To compare ensembles of Scrabble graphs from different languages, we require data. To this end, we use the Scrabble AI and analysis tool Quackle \cite{olaughlin_quackle_2019} to self-play and generate Scrabble graphs. The software allows us to define the lexicon and set of tiles at Quackle's disposal. In this section we give more details about the lexica and tile distributions used by Quackle, as well as the way it generates Scrabble games.  

\paragraph{Lexica and letter distributions} We use five languages included in the Quackle distribution: English, French, Greek, Norwegian, and Turkish. For English tournament Scrabble, the official word authority in North America is the NASPA Word List (NWL) \cite{naspa_nwl2023}, which we use here; its British counterpart is Collins Scrabble Words (CSW) \cite{collins_csw2024}. The dictionaries for the remaining languages are derived from a combination of official and open-source word lists. To enable Quackle to use the dictionaries, they must contain explicit enumeration of all possible conjugations and tenses. The Turkish dictionary included in Quackle is not representative of the full extent of the Turkish lexicon: Turkish is an agglutinative language, with many words formed by stringing together word parts (or morphemes). The word list in the Quackle distribution does not exhaust all permissible agglutinations that would typically be allowed in Scrabble: Nevertheless, we include Quackle Turkish here to observe the effects of a comparatively short dictionary on the structure of the Scrabble graphs. 

A crucial difference between lexica is the frequency at which particular letters are used: Official editions of Scrabble in different languages have different letter distributions, which means that the number of letters as well as their assigned value may differ. For example, in English editions of Scrabble, the letter `\verb+C+' occurs twice and is worth three points; in Norwegian editions, it occurs only once and is worth ten points. We use tile distributions consistent with official editions of Scrabble in each language \cite{scrabble_letter_distributions};  these, too, are included in the Quackle software distribution.

\paragraph{Quackle player modes}

Quackle has two types of computer-controlled `players' on offer: a \textit{Speedy Player} and a \textit{Championship Player}. The Speedy Player is the fastest player, basing its moves on a `static evaluation' only \cite{olaughlin_quackle_2019}. This evaluation ranks moves based on their score as well as the scoring potential of the unplayed tiles left on the rack. It is `static' in the sense that it does not look ahead or simulate different moves and how the opponent may counter them. While full mastery over the lexicon puts the Speedy Player at a significant advantage over most casual human players, it might make critical strategic errors; for example, by playing words that enable opponents to easily capture triple-word squares. The Championship Player overcomes this by first performing the static evaluation and then simulating opponent responses to a set of the most promising moves. While the Championship Player is therefore more positionally aware, it is also more computationally expensive. To enable simulation of ${\sim}10^5$ Scrabble games within a reasonable timescale, we use Quackle's Speedy Player. How the skill level or strategic insight of a (human or computer) player influences Scrabble graphs is an interesting open question. 

\section{Pairwise maximum entropy model predicts correlations and Scrabble graph geometry for all languages}

\begin{figure*}
\centering
\includegraphics[width=\textwidth]{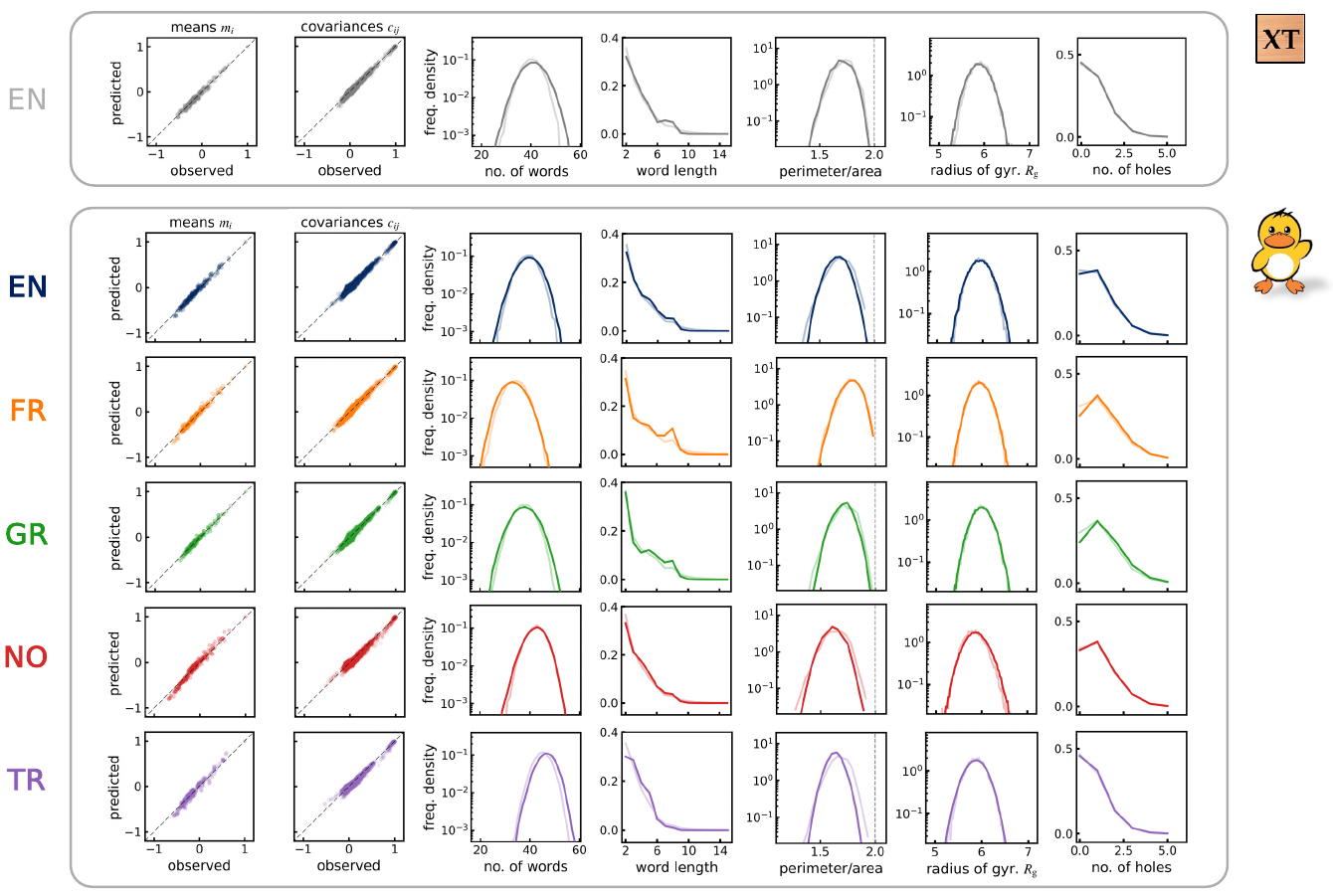}
\caption{We compare the predictions from the pairwise models $p(\boldsymbol{\sigma}\vert \ell)$, for all languages $\ell$, to the data. Shown is human English data (top row) and Quackle data for $\ell \in \{\text{English}, \text{ French}, \text{ Greek}, \text{ Norwegian}, \text{ Turkish}\}$ (bottom five rows). We show that the pseudo-likelihood method works and the models predict means and covariances as expected. We investigate several other observables to further probe the geometry of Scrabble graphs and test predictions from the model. From left to right: total number of words, word lengths, normalized perimeter, radius of gyration, and the number of holes in the Scrabble graph. Solid lines are data (from Cross-Tables or Quackle); lighter lines are predictions from the pairwise model.}\label{fig:languages_fit}
\end{figure*}

In this section, we verify that the PL method works for the ensembles of Scrabble graphs generated by Quackle: Indeed, the pairwise maximum entropy model predicts means and pairwise correlations well (Fig. \ref{fig:languages_fit}), satisfying the constraints $\langle \sigma_i\rangle = \langle \sigma_i \rangle^{\mathcal{D}}$ and $\langle \sigma_i \sigma_j \rangle = \langle \sigma_i \sigma_j \rangle^{\mathcal{D}}$ as intended by construction. 

The reproduction of pairwise statistics does not guarantee that the geometry of the Scrabble graphs has been captured; especially if there is intuition that structure may arise due to higher-order interactions between squares. To this end, we investigate several more observables: in the main text, we discussed the number of words on the board, the word lengths on the board and the normalized perimeter for human Scrabble tournament games; here, we add the radius of gyration and the number of holes (top row, Figure \ref{fig:languages_fit}). 

The radius of gyration $R_\text{g}$ measures the root-mean-square distance of $N$ tiles from the center of mass of the Scrabble graph, $R_\text{g}^2 = \sum_{i<j} (\boldsymbol{r}_i - \boldsymbol{r}_j)^2/N^2$, where $\boldsymbol{r}_i$ is the position of tile $i$. 
For human Scrabble graphs, it is approximately normally distributed around a mean $R_\mathrm{g}$ of approximately $6$ units. This mean radius of gyration is larger than $R_\mathrm{g}$ for a `disc' of tiles with area 99 $\text{units}^2$ (the maximum number of tiles), $R_\mathrm{g}^{\text{disk}}=3.9$ units, since Scrabble graphs contain more meandering structures and holes. The Scrabble graphs with the largest $R_\mathrm{g}$ contain more longer words at the periphery of the board. 
We find that the number of holes on the Scrabble board range between zero and six holes for human Scrabble, with a higher frequency of lower number of holes.

\begin{figure}
\includegraphics[width = 0.26\textwidth]{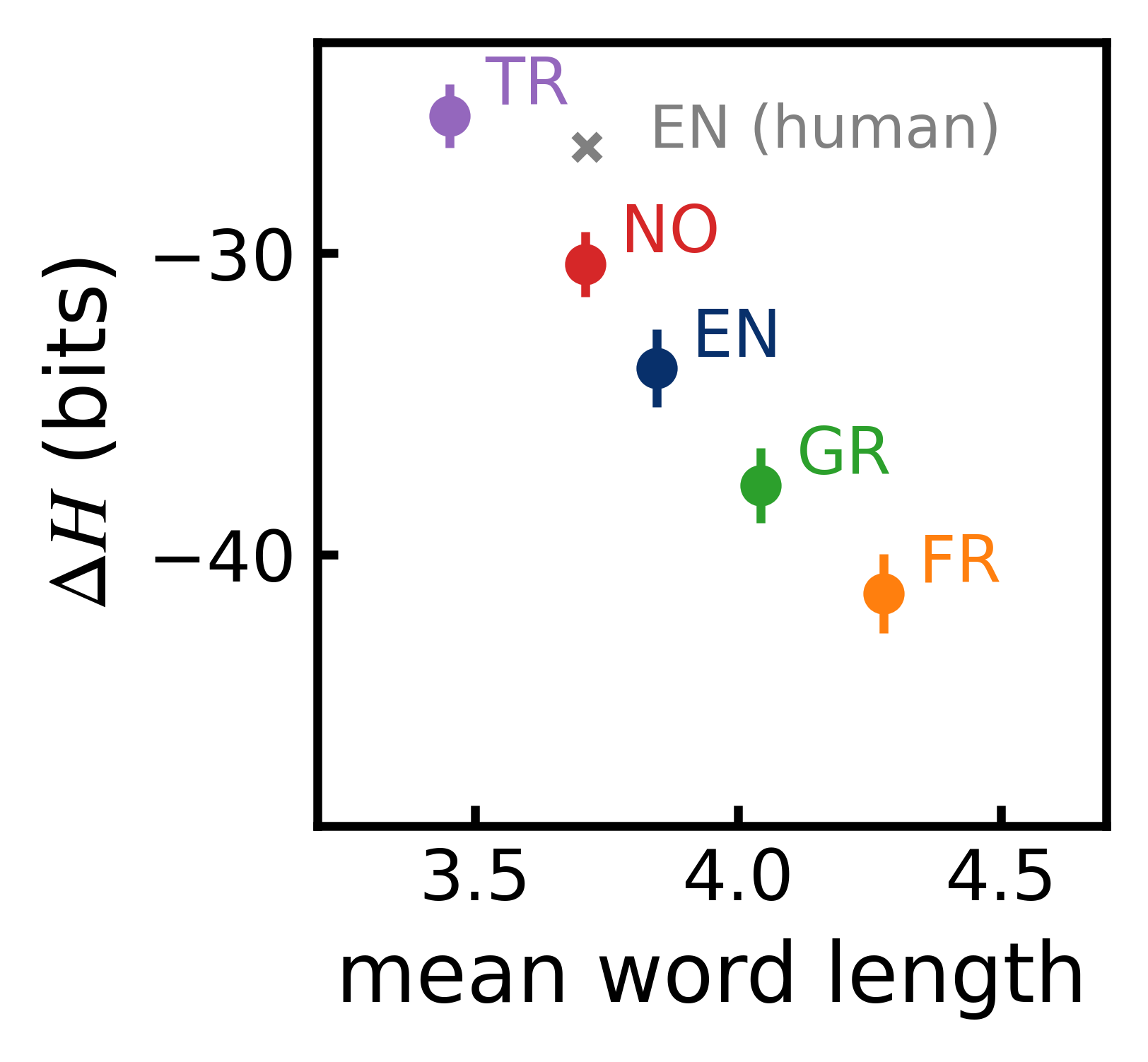}
\caption{Entropy decreases with mean length of words on the board across languages and player types. \label{fig:supp8}}
\end{figure}

We also show all these observables for the games generated by the Quackle AI tool (rows, Figure \ref{fig:languages_fit}). We find that in all cases, the pairwise model matches the data really well. The observables that seem most difficult to capture are the normalized perimeter and slight bumps in the word length distributions. We find that some of the features are correlated, for example, a higher number of words typically implies a shorter mean word length. However, there are exceptions, due to the intricate distribution of word lengths: for example, human English has a smaller mean number of words than Quackle Norwegian, but a 
similar
mean word length. Indeed, we find that the mean word length predicts 
Scrabble graph entropy 
slightly better (Fig. \ref{fig:supp8}). 

\section{Inferring language from Scrabble graphs using other methods}

\begin{figure}
\centering
\includegraphics[width=0.65\textwidth]{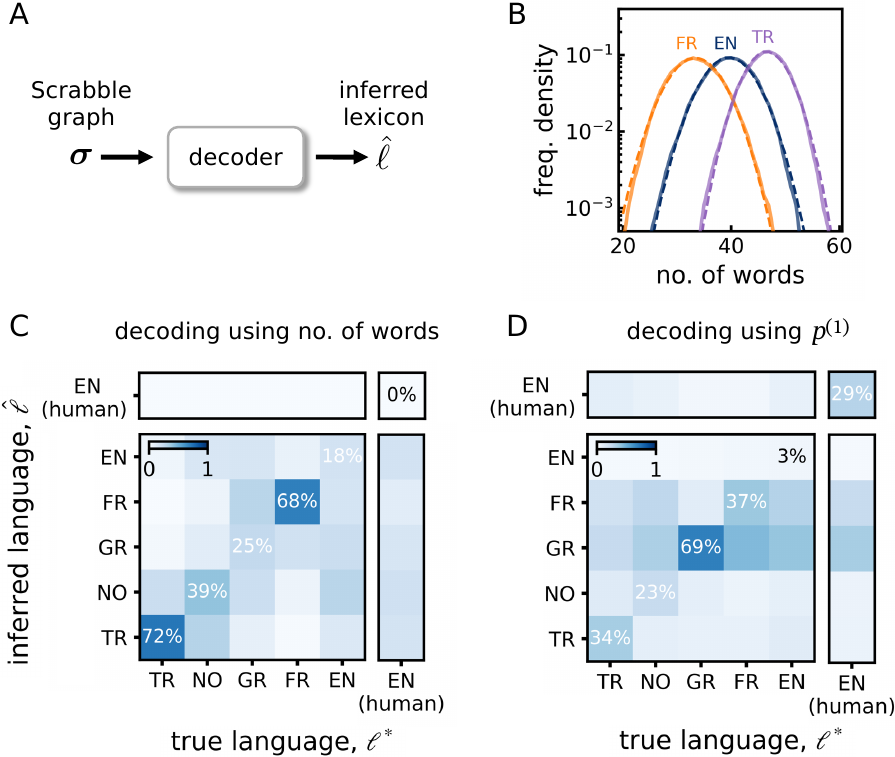}
\caption{(A) Schematic of the inference problem; we infer the lexicon $\ell$ given a particular Scrabble graph $\boldsymbol{\sigma}$. (B) Probability densities $p(w \vert \ell)$ for the total number of words $w$ on the board given a language $\ell$; three examples are shown (English, French, and Turkish; solid lines). We fit a Gaussian distribution to each density (broken lines). (C) Using the number of words to infer the language of the Scrabble graph: We make correct estimates roughly $40\%$ of the time. We note that human English never maximizes the \textit{a posteriori} probability in this simple setup, because its density $p(w \vert \ell)$ nowhere exceeds that of another language (not shown). (D) Decoding using $p^{(1)}$ leads to correct inference of the language approximately $30\%$ of the time. }\label{fig:decoding_p1}
\end{figure}

In the main text, we formulated a maximum entropy model $p^{(2)}(\boldsymbol{\sigma} \vert \ell)$ which we fit to Scrabble graph data for different languages $\ell$. Using Bayes' theorem---and thermodynamic integration to evaluate ratios of partition functions---we computed a posterior distribution $p(\ell \vert \boldsymbol{\sigma})$. From this posterior distribution, we can estimate the most likely language $\hat{\ell}$ that generated a particular Scrabble graph $\boldsymbol{\sigma}$. Remarkably, our simple, interpretable model constrained only by pairwise interactions is able to infer the correct language roughly $70\%$ of the time. 

We note that optimizing the performance of this Bayesian classifier (Fig. \ref{fig:decoding_p1}A) is not the goal of the manuscript: In principle, one could incorporate a large set of Scrabble graph features to improve decoding performance, in combination with more complicated network-based models. Rather, the purpose of our analysis is to identify how well low-order statistics already capture information about the underlying language. To this end, we compare the pairwise maximum entropy model to simpler baselines: a summary statistic, in this case the total number of words, and a model constrained only by the single-site occupation $p^{(1)}$. 

To infer the language $\ell$ of a Scrabble graph using the number of words $w$ on the board, we fit the distributions $p(w \vert \ell)$ for each language using a Gaussian distribution (Fig. \ref{fig:decoding_p1}B). When all languages are equiprobable, the posterior distribution is given by $p(\ell \vert w) = p(w \vert \ell)/\sum_{\ell'}p(w\vert \ell')$. As before, we infer the language $\hat{\ell}$ that maximizes the posterior distribution for all Scrabble graphs in a validation set; we keep track of our guesses in a confusion matrix (Fig. \ref{fig:decoding_p1}C). Performance is significantly worse than the pairwise model, with languages inferred correctly roughly $40\%$ of the time.
Similarly, the model $p^{(1)}$ also performs worse, with correct estimates roughly $30\%$ of the time (Fig. \ref{fig:decoding_p1}D). Computing the posterior for $p^{(1)}$, like for $p^{(2)}$, requires thermodynamic integration to evaluate ratios of partition functions.

\bibliography{scrabble.bib}